\documentclass[11pt]{article}
\pdfoutput=1
\usepackage{jcapmod}
\usepackage{booktabs}
\usepackage[english]{babel}
\usepackage{amsmath,amssymb,amsbsy,amstext, amsthm, simplewick}
\usepackage{graphicx}
\usepackage{amsfonts}
\usepackage{amssymb}
\usepackage{upgreek}
 \usepackage{exscale,relsize}
 \usepackage[makeroom]{cancel}
\usepackage{soul}

\RequirePackage{color}

\usepackage{colortbl}
\definecolor{rp}{cmyk}{0.2, 1, 0.6, 0}
\definecolor{green2}{cmyk}{0, 1, 0.5, 0}
\definecolor{lightgreen}{cmyk}{0.2, 0, 0.2, 0.2}
\definecolor{lightgray}{cmyk}{0.1,0.2,0,0.1}
\definecolor{lightgray2}{cmyk}{0.4,0.4,0,0.8}
\definecolor{black}{cmyk}{1.0,1.0,1.0,1.0}

\allowdisplaybreaks[1]

\usepackage{colortbl}
\definecolor{lightgreen}{cmyk}{0.2, 0, 0.2, 0.2}
\definecolor{lightgray}{cmyk}{0.1,0.2,0,0.1}
\definecolor{lightgray2}{cmyk}{0.1,0.1,0,0.1}

\makeatletter
\newlength{\apb@width}
\newcommand{\autoparbox}[2][c]{\settowidth{\apb@width}{#2}\parbox[#1]{\apb@width}{#2}}

\makeatother

\numberwithin{equation}{section}

\def\beq{\begin{equation}}
\def\eeq{\end{equation}}

\def\bea{\begin{eqnarray}}
\def\eea{\end{eqnarray}}

\def\d{{\rm d}}
\def\dd{{\rm d}}

\def\beq{\begin{equation}}
\def\eeq{\end{equation}}
\def\bea{\begin{eqnarray}}
\def\eea{\end{eqnarray}}

\def\d{{\rm d}}
\def\dd{{\rm d}}

\def\Mp{M_{\rm pl}}
\def\d{{\rm d}}

\def\b{{\vec b}}

\def\0{{\boldsymbol 0}}

\def\x{{\vec{x}}}
\def\y{{\vec{y}}}
\def\z{{\vec{z}}}

\def\k{{\vec{k}}}

\DeclareRobustCommand{\SkipTocEntry}[4]{}

\begin{document}

\begin{titlepage}

\setcounter{page}{1} \baselineskip=15.5pt \thispagestyle{empty}

\bigskip\

\vspace{1cm}
\begin{center}





{\fontsize{20}{24}\selectfont  \sffamily \bfseries  High-Scale Inflation and the Tensor Tilt}

\end{center}

\vspace{0.2cm}
\begin{center}
{\fontsize{13}{30}\selectfont  Daniel Baumann, Hayden Lee, and Guilherme L. Pimentel} 
\end{center}

\begin{center}

\vskip 8pt
\textsl{Department of Applied Mathematics and Theoretical Physics,\\ 
Cambridge University, Cambridge, CB3 0WA, UK}
\vskip 7pt

\end{center}

\vspace{1.2cm}
\hrule \vspace{0.3cm}
\noindent {\sffamily \bfseries Abstract} \\[0.1cm]
In this paper, we explore a novel observational signature of gravitational corrections during slow-roll inflation. We study the coupling of the inflaton field to higher-curvature tensors in models with a minimal breaking of conformal symmetry.  In that case, the most general correction to the tensor two-point function is captured by a coupling to the square of the Weyl tensor. We show that these scenarios lead to a correction to the tilt of the tensor power spectrum and hence a violation of the tensor consistency condition.  We arrive at the same conclusion through an analysis in conformal perturbation theory.  

\vskip 10pt
\hrule
\vskip 10pt

\vspace{0.6cm}
 \end{titlepage}

\tableofcontents

\newpage

\section{Introduction}

String theory predicts higher-curvature corrections to the gravitational effective action.  If the inflationary scale is sufficiently high, these corrections may be observable. 
In fact, the imprint of these effects on the primordial perturbations may be a rare observational window on inflationary models without a large hierarchy between the Hubble scale and the string scale.  These models are hard to analyze, since, unlike more conventional inflationary theories, they cannot be organized as an expansion in a ratio of energy scales.  To make robust statements therefore requires identifying observables which are protected by symmetries.  In this paper, we will consider inflationary models whose predictions are controlled by the weakly broken conformal symmetry of the quasi-de Sitter background.  Our approach is similar in spirit to that of~\cite{Maldacena:2011nz} and \cite{Arkani-Hamed:2015bza}  (see also~\cite{Antoniadis:2011ib, Creminelli:2011mw, Mata:2012bx, Ghosh:2014kba, Kundu:2014gxa, McFadden:2009fg, McFadden:2010vh, McFadden:2011kk, Bzowski:2011ab, Bzowski:2012ih, McFadden:2013ria, Kundu:2015xta}), where conformal symmetry was used to constrain the three-point functions for tensors and scalars, respectively.  In particular, in~\cite{Maldacena:2011nz}, it was shown that higher-curvature corrections give rise to a new structure in the graviton three-point function.  While the analysis of the tensor three-point function is particularly clean and model-insensitive, it is also hard to verify in observations, since tensor non-Gaussianities are likely to be very small.  Here, we will discuss a potentially larger signature in the graviton two-point function. 

\vskip 4pt
The tensor power spectrum is characterized by an amplitude (or the tensor-to-scalar ratio,~$r$) and a tilt ($n_t$).  In single-field slow-roll inflation, minimally coupled to Einstein gravity, these two parameters are related by a consistency condition~\cite{Lidsey:1995np}, $r = -8 n_t$.  We will show that the leading higher-curvature corrections to the gravitational action lead to a violation of this consistency condition.\footnote{Related observations have appeared in~\cite{Kaloper:2002uj}.} We will arrive at this conclusion from two different perspectives: 
\begin{itemize}
\item First, we consider a small breaking of de Sitter symmetry in the inflationary action, controlled by the slow-roll parameter $\varepsilon \equiv - \dot H/H^2$.  In our model, conformal symmetry is broken by the inflaton potential and a
coupling to the square of the Weyl tensor. These terms have been considered in the literature before, both from the perspective of an effective field theory~\cite{Weinberg:2008hq} and in string theory~\cite{zwiebach1985curvature} (see also~\cite{Brigante:2007nu,Kats:2007mq, Buchel:2008vz}).  
We argue that this simple action reproduces the most general corrections to the tensor two-point function at leading order in the breaking of the conformal symmetry. While the tensor-to-scalar ratio is of order $\varepsilon$, the tensor tilt gets a correction of order~$\sqrt{\varepsilon}\hskip 1pt (H/M)^2$, where $M$ is the scale suppressing the higher-curvature corrections. If the scale $M$ is close to the Hubble scale, this correction is the dominant effect.\footnote{Corrections due to higher-derivative interactions of the inflaton, such as $(\partial \phi)^4/\Lambda^4$, will be suppressed by a larger scale $\Lambda$. The hierarchy $\Lambda \gg M$ is made technically natural by the underlying conformal symmetry.} 

\item Second, we analyze the same problem from the point of view of the wavefunction of the universe.  The coefficients of the wavefunction are constrained by the de Sitter isometries, and can be interpreted as correlation functions of a putative conformal field theory~(CFT)~\cite{Strominger:2001pn, Maldacena:2002vr}.  To describe a realistic cosmology, we break the conformal symmetry by introducing a marginally relevant deformation of the CFT.  If the deformation is small, then it can be treated in conformal perturbation theory \cite{Zamolodchikov:1987ti}.
The slow-roll parameter $\varepsilon$ is related to the coupling and the dimension of the operator that deforms the CFT~\cite{Larsen:2003pf}. We use conformal perturbation theory as a robust proof of our claim that the tensor-to-scalar ratio~$r$ and the tensor tilt~$n_t$ are non-zero at different orders in $\sqrt{\varepsilon}$.   
The advantage of this approach is that it relies mostly on symmetries, and allows us to be agnostic about the details of the effective action of the theory.
\end{itemize}

\vskip 10pt
\noindent
The outline of the paper is as follows.
In Section~\ref{sec:bulk}, we introduce a simple action in which higher-curvature corrections lead to a violation of the tensor consistency condition. We emphasize that the effect arises at leading order in the breaking of the de Sitter isometries.  In Section~\ref{sec:boundary}, we confirm this conclusion with an analysis of the stress tensor two-point function in a perturbed conformal field theory. Our conclusions are summarized in Section~\ref{sec:conclusions}.  Two appendices contain technical details.
In Appendix~\ref{app:cftbreaking}, we make a few remarks on the breaking of conformal symmetry in theories with a nontrivial sound speed. In Appendix~\ref{app:Einstein}, we review the  argument of \cite{Creminelli:2014wna}, showing that a nontrivial tensor spectrum can be mapped to a nontrivial scalar spectrum by a disformal transformation~\cite{Bekenstein:1992pj}.

\subsubsection*{Notation and Conventions}

We will use natural units, $c = \hbar =1$, with reduced Planck mass  $\Mp^2 \equiv 1/8\pi G$. Our metric signature is $(-+++)$.  Latin indices denote spatial coordinates (e.g.~$x_i$, $k_i$), while Greek indices stand for spacetime coordinates.  We will use both physical time $t$ and conformal time $\eta$. 
Overdots and primes will denote derivatives with respect to $t$ and $\eta$, respectively. 
The dimensionless power spectrum of a Fourier mode $f_\k$ is defined as
\beq
\Delta_f^2(k) \equiv \frac{k^3}{2\pi^2}\langle f_\k f_{-\k}\rangle'\ ,
\eeq
where the prime on the correlation function indicates that the overall momentum-conserving delta function is being dropped.

\section{Tensors beyond Einstein Gravity}
\label{sec:bulk}

In this section, we motivate a scenario in which the inflaton breaks the isometries of de Sitter space by a minimal amount, and analyze the consequences for the scalar and tensor two-point functions. 
We show that a coupling to the Weyl tensor leads to a violation of the tensor consistency condition.

\subsection{Weakly Broken Conformal Symmetry}

\begin{figure}[h!]
\centering
\includegraphics[scale=1.1]{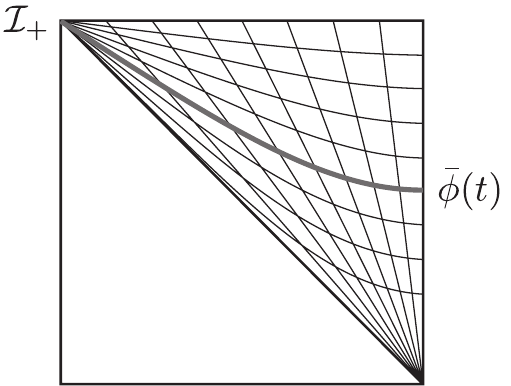}
\caption{The time-dependent inflaton vev, $\bar \phi(t)$, introduces a preferred time slicing of de Sitter space.} 
\label{fig:Penrose}
\end{figure}

De Sitter space is a solution to Einstein gravity with a positive cosmological constant
\beq
S_{dS}=\frac{\Mp^2}{2}\int \dd^4x\sqrt{-g}\Big(R -2\Lambda \Big)\ .
\eeq
 In flat slicing, the line element is given by
\beq\label{dsmetric}
\d s^2 = -\d t^2+e^{2Ht} \d\vec{x}\hskip 1pt{}^2
=\frac{-\d\eta^2+\d\vec{x}\hskip 1pt{}^2}{(H\eta)^2}\ ,\quad \text{with} \quad 3 H^2 \equiv \Lambda\ .
\eeq
Inspection of \eqref{dsmetric} reveals the isometries of the de Sitter spacetime: in addition to spatial rotations and  translations, these include a dilatation 
and three special conformal transformations 
\begin{align}
{\rm D}:& \qquad \eta \,\to\, \eta \hskip 1pt (1+ \lambda) \ , \quad \hspace{.9cm}  \vec{x} \,\to\, \vec{x} \hskip 1pt (1+\lambda)\ , \\
{\rm SCT}:& \qquad \eta \,\to\, \eta(1 - 2 {\vec b}\hskip 1pt.\hskip 1pt {\vec x}\hskip 1pt)\ , \qquad \vec{x} \,\to\, \vec{x}-2(\hskip 1pt \vec{b}\hskip 1pt.\hskip 1pt \vec{x}\hskip 1pt) \hskip 2pt \vec{x}+(\vec{x}\hskip 1pt{}^2-\eta^2)\hskip 1pt \vec{b}\ ,
\end{align}
where $\lambda$ and $\vec b$ are infinitesimal parameters.
At late times, $\eta \to 0$, these isometries act as conformal transformations on the spacelike boundary~${\cal I}_+$ (see figure \ref{fig:Penrose}).
We will return to this point of view in Section~\ref{sec:boundary}.

\vskip 4pt
In a realistic inflationary model, the de Sitter symmetries need to be broken. For this purpose, we introduce the dynamical inflaton field $\phi$. We give it a potential, $\Mp^2 \Lambda \to V(\phi)$, so that the field acquires a time-dependent expectation value $\phi=\bar\phi(t)$.
This provides a natural clock measuring the time to the end of inflation.
The expansion rate is now time dependent, $H \to H(t)$, and related to the evolution of the inflaton by the Einstein equations~\cite{Baumann:2009ds} 
\begin{align}
\dot{\bar \phi}\hskip 1pt{}^2 &=-2 \Mp^2 \dot H \ ,\\[4pt]
V(\bar\phi) &= \Mp^2\big(3H^2+\dot H\big)\ .
\end{align}
The inflationary slow-roll parameter can then be written as
\beq\label{epb}
\varepsilon = \frac{\frac{1}{2}\dot{\bar \phi}^2}{\Mp^2 H^2}\ .
\eeq
The size of $\varepsilon$ controls the breaking of the conformal symmetries of de Sitter, with the symmetries being restored in the limit $\varepsilon \to 0$.

\vskip 4pt
We will assume that inflaton self-interactions are suppressed by a relatively large mass scale, $ \Lambda^2 \gg \dot{\bar\phi}$, while gravitational interactions are controlled by a lower scale, $\dot{\bar\phi} > M^2 \gtrsim H^2$.  
The hierarchy $\Lambda \gg M$ is stabilized by the underlying conformal symmetry.
In that case, the leading breaking of the conformal symmetry comes from the inflaton potential, while higher-derivative interactions, like $(\partial \phi)^4/\Lambda^4$, will be suppressed\footnote{If the scale controlling inflaton interactions is smaller than $\dot{\bar \phi}$, then the power counting of the EFT changes significantly~\cite{Cheung:2007st}.  In this limit, inflaton fluctuations can propagate with a nontrivial sound speed, $c_s \ll 1$.  In Appendix~\ref{app:cftbreaking}, we show that a small sound speed induces a much stronger breaking of conformal symmetry than we wish to consider in this paper. Conversely, if we demand that the conformal symmetry is only broken by effects of order $\varepsilon$, then these symmetry-breaking operators have to be highly suppressed. } by powers of $\dot{\bar \phi}/\Lambda^2 \ll H^2/M^2$.  In addition, we may have functions of $\phi$ coupled to curvature tensors.  These couplings were discussed systematically by Weinberg in~\cite{Weinberg:2008hq}. Like Weinberg, we consider these terms to be perturbative corrections to the Einstein-Hilbert action. This ensures that any ghost instabilities are moved outside the regime of validity of the effective theory.
Using the field equations of the leading terms in the action, all inflaton-curvature couplings can be written in terms of couplings to the Weyl tensor\hskip 1pt\footnote{Couplings to $R^2$ and $R^{\mu \nu} R_{\mu \nu}$ can be removed by using the leading-order equation of motion~\cite{Weinberg:2008hq}, $\Mp^2 R_{\mu \nu} = - \partial_\mu \phi \partial_\nu \phi - V(\phi) g_{\mu \nu}$. This leads to corrections to the inflaton action, which, as before, we assume are constrained by conformal symmetry.}
\beq
W_{\mu \nu \hskip 1pt \rho \sigma} \equiv R_{\mu \nu \hskip 1pt \rho \sigma} - \frac{1}{2} \left(g_{\mu \rho} R_{\nu \sigma} -g_{\mu \sigma} R_{\nu \rho} - g_{\nu \rho} R_{\mu \sigma} + g_{\nu \sigma} R_{\mu \rho}\right) + \frac{R}{6} (g_{\mu \rho} g_{\nu \sigma} - g_{\nu \rho} g_{\mu \sigma})\ .
\eeq
To study the tensor two-point function, we only need to consider the couplings to the square of the Weyl tensor,
\beq
W^2 \equiv W^{\mu \nu\hskip 1pt \rho \sigma} W_{\mu \nu \hskip 1pt \rho \sigma} = R^{\mu \nu\hskip 1pt \rho \sigma} R_{\mu \nu \hskip 1pt \rho \sigma} - 2 R^{\mu \nu}R_{\mu \nu} +\frac{1}{3} R^2\ ,
\eeq
 and to the parity-violating term $W \widetilde W \equiv (\sqrt{-g}\hskip 1pt)^{-1} \hskip 1pt \epsilon^{\mu \nu\hskip 1pt \rho \sigma} W_{\mu \nu}{}^{\kappa \lambda}W_{\rho \sigma \hskip 1pt \kappa \lambda}$.  
Higher powers of the Weyl tensor will contribute to higher-point correlation functions, and are not relevant for the considerations of this paper.
We will therefore study the following action
\beq
S=\int \dd^4x\, \big({\cal L}_\phi+{\cal L}_g\big)\ , \quad \ \text{with} \quad \ \begin{aligned}{\cal L}_\phi &\ = \ \sqrt{-g}\, \left[- \frac{1}{2}(\partial \phi)^2 - V(\phi)\right] \ ,\\[4pt]
 {\cal L}_g &\ =\  \sqrt{-g}\,\frac{\Mp^2}{2} \left[ R + f(\phi) \frac{W^2}{M^2} + h(\phi) \frac{W \widetilde W}{M^2}\right]\ .
 \end{aligned} \label{eq:action}
\eeq
Notice that we have factored out the scale $\Mp^2$ in ${\cal L}_g$. This is consistent with the structure expected in string effective actions~\cite{Baumann:2014nda}, with $M$ playing the role of  the string scale or the Kaluza-Klein scale.
Since the Weyl tensor vanishes for any homogeneous FRW metric, the background slow-roll solution is still determined by the Einstein-Hilbert part of ${\cal L}_g$.  

\vskip 4pt
The effects of the parity-violating term $h(\phi) W \widetilde W$ have been studied in~\cite{Lue:1998mq, Alexander:2004wk, Contaldi:2008yz, Takahashi:2009wc}. 
Since $W \widetilde W$ is a total derivative, this term vanishes if $h(\phi)$ is a constant. 
The correction to the tensor two-point function therefore comes from the field-dependent variation of $h(\phi)$.  This leads to a difference in the amplitudes of the two chiralities of the tensor modes of order $\sqrt{\varepsilon} \hskip 1pt H^2/M^2$.  
In this paper, we will be interested in the effects of the coupling $f(\phi) W^2$. We will show that the field-dependent variation\footnote{For constant $f(\phi)$, the Weyl-squared term, $W^2$, can be put into the Gauss-Bonnet  form, $R_{\mu \nu \hskip 1pt \rho \sigma}^2 - 4 R_{\mu \nu}^2 +  R^2$, via the field redefinition $g_{\mu\nu}\to g_{\mu\nu}+ f\Mp^{-2} (-2 \hskip 1pt R_{\mu\nu}+\frac{5}{3}\hskip 1pt  g_{\mu\nu} R)$~\cite{Deser:1986xr}.  Since the Gauss-Bonnet term is a total derivative, it only contributes a boundary term.  However, the field redefinition also changes the inflaton kinetic term and the normalization of the Einstein-Hilbert action, so the constant part of the function $f(\phi)$ is still physical.} of $f(\phi)$ induces a sound speed for tensor fluctuations and a contribution to the tensor tilt of order~$\sqrt{\varepsilon} \hskip 1pt H^2/M^2$.

\vskip 4pt
Interestingly, the $ f(\phi) W^2$ term in \eqref{eq:action} is similar to a term in the effective action of the original Starobinsky model~\cite{Starobinsky:1980te}.  In these models, inflation is driven by a large number of conformally coupled fields whose stress tensor is induced by the conformal anomaly~\cite{Duff:1993wm}, $\langle T^\mu_\mu \rangle \supset c \hskip 1pt W^2$. The effective action that reproduces the conformal anomaly~\cite{Riegert:1984kt} includes the Weyl-squared term.  It is not hard to imagine that variations of the model could contain a term of the form $ f(\phi) W^2$. For example, such a term arises if one introduces the dilaton. It would be interesting to make this connection more precise~\cite{future}.\footnote{We thank Juan Maldacena for this suggestion.}

\subsection{Violation of the Consistency Condition}

We now compute the scalar and tensor power spectra resulting from the action~\eqref{eq:action}. 
We use the standard ADM decomposition of the metric~\cite{Arnowitt:1962hi}
\beq
\d s^2=-N^2 \d t^2+g_{ij}\big(N^i \d t + \d x^i \big)\big(N^j \d t+\d x^j\big)\ .
\eeq
In comoving gauge, the inflaton is unperturbed, $ \phi = \bar \phi(t)$, and the spatial metric can be written as
 \beq
g_{ij} = a^2 e^{2 \zeta} \left(\delta_{ij}+ \gamma_{ij}\right)\, , \label{equ:comoving}
 \eeq
 where $\gamma_{ij}$ is a transverse and traceless tensor.
At leading order, the curvature perturbation $\zeta$ and the tensor mode $\gamma_{ij}$ decouple and can be treated separately.

 \subsubsection*{Scalars}
 
 We first consider the spectrum of the curvature perturbation $\zeta$.  Since $W \widetilde W$ vanishes for scalar fluctuations, only $f(\phi) W^2$ contributes.  At linear order in $\zeta$, the non-zero components of the Weyl tensor are\hskip 1pt\footnote{To arrive at (\ref{equ:WeylZeta}), we have used the linearized solutions to the Einstein constraint equations~\cite{Maldacena:2002vr}: $\delta N =\dot \zeta/H$ and $\partial_i N^i = \varepsilon \dot \zeta - a^{-2}\nabla^2 \zeta/H $.}
\begin{equation}
\begin{aligned}
W^{0i}{}_{0j}&= \frac{1}{2}\left(\partial_i\partial_j-\frac{\delta_{ij}}{3}\nabla^2 \right)\left(\frac{\varepsilon \hskip 1pt \zeta}{a^2} + \frac{1}{\nabla^2}\frac{d}{dt}(a \varepsilon \hskip 1pt \dot\zeta)\right)\equiv F_{ij}\ , \\[4pt]
W^{ij}{}_{kl}&=\delta_{ik}F_{jl}+\delta_{jl}F_{ik}-\delta_{il}F_{jk}-\delta_{jk}F_{il}\ ,
\end{aligned}
\label{equ:WeylZeta}
\end{equation}
and hence $W^2 = 8 F_{ij}^2$.  To eliminate  terms with second-order time derivatives in the Weyl tensor, we use the leading-order equation of motion
\beq
\ddot{\zeta}+\frac{d}{dt}\log(a^3 \varepsilon) \,\dot\zeta-\frac{\nabla^2}{a^2}\zeta=0\ . \label{equ:EOMsc}
\eeq
The quadratic action can then be written in the following form 
 \begin{align}
 \frac{{\cal L}_\zeta}{\Mp^2} &\ = \  a^3\,  \frac{\varepsilon}{c_s^2} \left(  \dot \zeta^2 - \frac{c_s^2}{a^2} (\vec{\nabla} \zeta)^2\right)  \ ,  
 \end{align}
 where we have introduced the sound speed
 \beq
 \frac{1}{c_s^2} - 1 \,\equiv \,  \frac{8}{3} \varepsilon f(\bar \phi) \frac{H^2}{M^2} \ .
 \eeq
If we had kept the couplings of the inflaton to $R^2$ and $R_{\mu \nu}^2$, we would have found additional corrections to $c_s$ of the same order. 
Since the deviation from $c_s =1$ is suppressed by a factor of $\varepsilon \ll 1$, it will not play a significant role for the rest of this paper.  For simplicity, we will therefore take $c_s \approx 1$, and write the power spectrum of $\zeta$ in the standard slow-roll form
\beq
\Delta_\zeta^2 \approx \frac{1}{8\pi^2}  \frac{1}{\varepsilon}\frac{H^2}{\Mp^2}\ ,
\eeq
where the right-hand side is evaluated at horizon crossing, $k=aH$.
We conclude that the coupling to the Weyl tensor has very little effect on the scalar power spectrum, and its main effect is a correction to the tensors.

 \subsubsection*{Tensors}
 
The linearized equation of motion for tensor fluctuations in Einstein gravity is 
\beq
\ddot{\gamma}_{ij}+3H \dot\gamma_{ij}-\frac{\nabla^2}{a^2}\gamma_{ij}=0\ . \label{equ:EOMg}
\eeq
We will use this to simplify some of the perturbative corrections to the quadratic action for $\gamma$. 
At linear order in $\gamma$, the components of the Weyl tensor are
\begin{align}
W^{0i}{}_{0j}&=\frac{1}{4}\left(\ddot\gamma_{ij}+H\dot\gamma_{ij}+\frac{\nabla^2}{a^2}\gamma_{ij} \right)\ ,\nonumber\\
W^{0i}{}_{jk}&=\frac{1}{2 a}\Big(\dot\gamma_{ik,j}-\dot\gamma_{ij,k}\Big) \ , \nonumber\\
W^{jk}{}_{0i}&=\frac{1}{2 a}\Big(\dot \gamma_{ij,k}-\dot\gamma_{ik,j}\Big)\ , \\
W^{ij}{}_{kl}&=\frac{1}{2}\left\{ \frac{1}{a^2}\Big(\gamma_{il,jk}+\gamma_{jk,il}-\gamma_{ik,jl}-\gamma_{jl,ik} \Big) \right.\nonumber \\
& \qquad \ +\frac{1}{2}\left[\delta_{il}\left(\ddot{\gamma}_{jk}+H\dot\gamma_{jk}-\frac{\nabla^2}{a^2}\gamma_{jk}\right)   \,+\,\delta_{jk}\left(\ddot{\gamma}_{il}+H\dot\gamma_{il}-\frac{\nabla^2}{a^2}\gamma_{il}\right) \right. \nonumber \\
& \qquad \ \qquad \ \left. \left.- \delta_{ik}\left(\ddot{\gamma}_{jl}+H\dot\gamma_{jl}-\frac{\nabla^2}{a^2}\gamma_{jl}\right)-\delta_{jl}\left(\ddot{\gamma}_{ik}+H\dot\gamma_{ik}-\frac{\nabla^2}{a^2}\gamma_{ik}\right)\right] \right\}\ .\nonumber
\end{align}
Substituting this into $W^2$, we get
\begin{align}
W^{\mu\nu}{}_{\rho\sigma}W^{\rho\sigma}{}_{\mu\nu} &\,=\, 4W^{0i}{}_{0j}W^{0j}{}_{0i}+4W^{0i}{}_{jk}W^{jk}{}_{0i}+W^{ij}{}_{kl}W^{kl}{}_{ij} \ , \nonumber \\[2pt]
&\,=\, 2 \left(\ddot\gamma_{ij}+H\dot\gamma_{ij}+\frac{\nabla^2}{a^2}\gamma_{ij}\right)^2+ 4 \hskip 1pt \dot\gamma_{ij}\frac{\nabla^2}{a^2}\dot\gamma_{ij}\ .
\end{align}
Using (\ref{equ:EOMg}), this can be brought into the form of eq.~(21) in~\cite{Weinberg:2008hq}.  After a few integrations by parts, we obtain 
 \begin{align}
\frac{ {\cal L}_\gamma}{\Mp^2} &\ = \ \frac{a^3}{8} \frac{1}{c_t^2}\left(\dot \gamma_{ij}^2 - \frac{c_t^2}{a^2}(\vec{\nabla} \gamma_{ij})^2 \right)\nonumber \\[4pt]
&\ \ \ \ \ -2 \frac{a^3}{M^2} \,f(\bar\phi) \left[\gamma_{ij}\frac{\nabla^2}{a^2}\left(\ddot\gamma_{ij}+3H\dot\gamma_{ij}-\frac{\nabla^2}{a^2}\gamma_{ij}\right) \right] \nonumber \\[4pt]
 &\ \ \ \ \ - 2\frac{a^3}{M^2}\, \frac{d f(\bar\phi)}{d t} \gamma_{ij} \frac{\nabla^2}{a^2} \dot\gamma_{ij} - \frac{4}{M^2}\, \frac{d h(\bar\phi)}{d t} \, \epsilon^{ijk0} \, \gamma_{il} \partial_j \nabla^2 \gamma_{kl} \ . \label{equ:Lgamma}
 \end{align}
The second line in (\ref{equ:Lgamma}) vanishes after using the equation of motion (\ref{equ:EOMg}) once more. The last line is proportional to $\dot {\bar \phi}$ and hence is suppressed in the slow-roll limit.\footnote{The last term in (\ref{equ:Lgamma}), although slow-roll suppressed, is phenomenologically interesting because it leads to chiral gravitational waves~\cite{Lue:1998mq, Alexander:2004wk, Contaldi:2008yz, Takahashi:2009wc}. Incidentally, the size of the chiral splitting, $\sqrt{\varepsilon} \hskip 1pt H^2/M^2$, is of the same order as the correction to the tensor tilt that we will get from the rest of the action.} This leaves the first line, which is the quadratic action for tensors with a nontrivial sound speed  
 \beq
 \frac{1}{c_t^2} - 1 \,\equiv \,  8 f(\bar \phi) \frac{H^2}{M^2} \ .  \label{equ:ct}
  \eeq
In~\cite{Creminelli:2014wna}, it was shown that a tensor sound speed can always be set to unity by a disformal transformation~\cite{Bekenstein:1992pj}, followed by a Weyl rescaling to take the action to Einstein frame.  These two metric transformations trade the nontrivial tensor sound speed for a scalar sound speed, $\tilde c_s$, and a modified Hubble rate, $\tilde H(\tilde t\hskip 1pt)$.  In the new frame, the tensor spectrum takes the standard form, $\Delta_\gamma^2 \propto \tilde H^2/\Mp^2$, but the scalar spectrum is modified. Of course, predictions for observables are frame-independent~\cite{Creminelli:2014wna, Tsujikawa:2014uza}, so the choice of frame is simply a matter of convenience. In particular, the violation of the consistency condition that we will find is a frame-independent conclusion.\footnote{We thank Paolo Creminelli and Filippo Vernizzi for a discussion of these issues.}   We leave the details to Appendix~\ref{app:Einstein}, but one point is worth emphasizing here. The violation of the consistency condition in the new frame is not the same as that found in $P(X)$-theories~\cite{Chen:2006nt}. Rather, it is the time derivative of $\tilde c_s$ that modifies the tensor-to-scalar ratio. 

The tensor sound speed leads to a simple rescaling of the standard tensor power spectrum.
Summing over the two graviton polarizations, we obtain
\beq
\Delta_\gamma^2 = \frac{2}{\pi^2}\frac{H^2}{\Mp^2}  \frac{1}{c_t}\ , \label{equ:Dg}
\eeq
where the right-hand side is evaluated at $c_t k = aH$.
 If the inflaton-Weyl coupling is a small correction to the leading gravitational action---as we are assuming in order to avoid propagating ghost degrees of freedom---then $c_t$ can't deviate much from unity.
 The main effect is not the size of $c_t$, but its time dependence.\footnote{The fact that $c_t$ is never allowed to deviate too far from unity puts a constraint on the time dependence of $c_t$, and hence on the function $f(\phi)$ in~(\ref{equ:ct}). }  In particular, the tensor-to-scalar ratio approximately still takes the form predicted by standard slow-roll inflation
\beq
r \,\equiv\, \frac{\Delta_\gamma^2}{\Delta_\zeta^2} \,=\, \frac{16 \varepsilon}{c_t} \approx 16 \varepsilon \ .  \label{equ:r} 
\eeq
However, the tensor tilt can still receive an important correction due to the time dependence of~$c_t$.
 Crucially, the evolution of $c_t(t)$ is coupled to that of $\bar \phi(t)$.  This will induce a tilt of the tensor spectrum proportional to $\dot{\bar \phi} \propto \sqrt{\varepsilon}$. To see this, let us define a slow-variation parameter for the tensor sound speed, $\varepsilon_t$, and express it in terms of the slow-roll parameter $\varepsilon$:
 \begin{align}
\varepsilon_t \equiv \frac{\dot c_t}{Hc_t} &\,= \,  \mp\hskip 1pt 4 \hskip 1pt c_t^2 \hskip 1pt b\hskip 1pt \sqrt{2\varepsilon}\hskip 1pt \frac{H^2}{M^2} \hskip 1pt + \hskip 1pt (1-c_t^2)\hskip 1pt \varepsilon \ , 
\end{align}
where $b \equiv \Mp f'$ is a dimensionless constant. Taking the scale of variation of $f(\phi)$ to be of order~$\Lambda$, we get $b \sim \Mp/\Lambda$, which may be large if $\Lambda \ll \Mp$.  The fractional change of (\ref{equ:Dg}) per Hubble time then determines the tensor tilt 
\beq
n_t \, \equiv \, \frac{d \ln \Delta_\gamma^2}{d \ln k} \, =\, -2 \varepsilon - \varepsilon_t\,\approx\, -2\varepsilon \,\pm\, 4b \hskip 1pt \sqrt{2\varepsilon} \hskip 1pt \frac{H^2}{M^2} \ , \label{equ:nt}
\eeq
where we have ignored a small shift in the coefficient of the standard contribution, $-2\varepsilon$.  
Notice that the correction has undetermined sign, so it seems to allow a blue tilt for the tensor spectrum, even without a violation of the null energy condition (NEC).\footnote{In the Einstein frame, with $\tilde c_t = 1$, a blue tilt still corresponds to a violation of the NEC, $\dot{\tilde H} > 0$, but without inducing the gradient instability that this usually implies~\cite{Creminelli:2006xe, Creminelli:2014wna}.} The second term in (\ref{equ:nt}) leads to a modification of the tensor consistency condition
\beq
\boxed{ - \frac{8 \hskip 1pt n_t}{r} \,=\, 1 \mp \frac{4 b}{\sqrt{2\varepsilon}} \frac{H^2}{M^2} }\ .\label{eq:consistcond}
\eeq
We see that the violation of the relation $n_t = -r/8$ is enhanced for small $\varepsilon$ (and large $b$), but suppressed by $H^2/M^2$. In the stringy regime of inflation, $H^2/M^2$ can be of order one\footnote{When $M$ is the string scale, the ratio $H/M$ is constrained by the fact that we require the Hagedorn temperature to remain above the de Sitter temperature in order to avoid a phase transition of the system~\cite{atick1988hagedorn}.} and our proposed modification of the tensor spectrum could be a significant effect. 

Testing the tensor consistency condition observationally is challenging (see~\cite{Caligiuri:2014sla, Dodelson:2014exa} for a recent discussion). Naturally, the observational prospects improve for large tensor-to-scalar ratio and if a large range of scales can be accessed (maybe with futuristic direct detection experiments~\cite{Smith:2005mm, Smith:2006xf,chongchitnan2006prospects}).  A blue tensor spectrum would be easier to detect.

\section{CFT Interpretation}
\label{sec:boundary}

The freeze-out of quantum fluctuations during inflation allows us to recast cosmological expectation values in terms of the `wavefunction of the universe', $\Psi[\zeta,\gamma]$. This wavefunction computes late-time expectation values of superhorizon fluctuations. The isometries of de Sitter space imply that the coefficients of the wavefunction can be interpreted as correlation functions of the stress tensor in a putative conformal field theory~\cite{Strominger:2001pn,Maldacena:2002vr}. The small breaking of conformal symmetry during inflation is modelled as a small deformation of the CFT, which can be treated perturbatively~\cite{Zamolodchikov:1987ti}.  In this section, we will show that this alternative point of view reproduces the results of the previous section.

\subsection{Wavefunction of the Universe}

The wavefunction of the universe can be computed by a saddle-point approximation, $\Psi \approx e^{i S_{cl}}$, where the action $S_{cl}$ is evaluated for a classical solution with certain Dirichlet boundary conditions~\cite{Maldacena:2002vr}. The result takes the following form
\beq
\Psi=e^{i S_{div}} e^{W_0[\zeta,\gamma]}\ ,\quad\ \text{with} \quad\ W_0=\frac{1}{2}\int \dd^3 k \left(\zeta_{\k}\hskip1pt\zeta_{-\k}\,\langle T_{\k} \hskip1pt T_{-\k} \rangle' +\sum_{s}\gamma^s_{\k}\hskip1pt \gamma^{s}_{-\k}\, \langle T^s_{\k}\hskip1pt T^{s}_{-\k}\rangle' \right) .
\eeq
 The local divergent piece, $e^{i S_{div}}$, is a pure phase factor, and thus drops out of expectation values.  The coefficient functions $\langle T_{\k} T_{-\k} \rangle'$ and $\langle T^s_{\k}T^{s}_{-\k}\rangle'$ may be interpreted as the correlation functions of the trace and the trace-free part of the stress tensor $T_{ij}$ of a dual field theory~\cite{Strominger:2001pn,Witten:2001kn,Maldacena:2002vr}.\footnote{The CFT that describes the de Sitter cosmology is not unitary and has some unusual features, mostly related to the spectrum of the dimensions of operators.  Our analysis will only use the mapping of the symmetries between the bulk and the boundary, and does not rely on a deeper meaning of dS/CFT.} The power spectra of $\zeta$ and $\gamma$ are then computed by a simple Gaussian integration 
\begin{align}
\label{eq:dictionary}
\langle \zeta_{\k} \hskip1pt \zeta_{-\k}\rangle' &\ =\  \int {\cal D}\zeta\,  \zeta_{\k} \hskip1pt \zeta_{-\k}\, |\Psi[\zeta]|^2\ \ \ \ \hskip 1pt =\ -\frac{1}{2 \hskip 1pt \text{Re}\langle T_{\k} \hskip1pt T_{-\k} \rangle' }\ ,\qquad \\ 
\langle \gamma^s_{\k} \hskip1pt \gamma^{s}_{-\k}\rangle' &\ =\ \int {\cal D}\gamma^s\,  \gamma_{\k}^s \hskip1pt \gamma_{-\k}^s\, |\Psi[\gamma^s]|^2 \ =\ - \frac{1}{2  \hskip 1pt \text{Re}\langle T^s_{\k}\hskip1pt T^s_{-\k} \rangle' }\ .
\end{align}
The diffeomorphism invariance of gravitational theories implies that the generators of coordinate transformations act as constraints on the wavefunction~\cite{DeWitt:1967yk}.
These constraint equations are the conformal Ward identities of the coefficient functions $\langle T_{\k} \hskip1pt T_{-\k} \rangle'$ and $\langle T^s_{\k}\hskip1pt T^s_{-\k} \rangle'$. In a CFT, these constraints imply 
\beq
\langle T_{\k} \hskip1pt T_{-\k}\rangle'=0\ ,  \quad\ \langle T^s_{\k}\hskip1pt T^s_{-\k}\rangle' = c_T k^3\ ,
\eeq 
where $c_T$ is the central charge. We see that there are no $\zeta$-fluctuations and the gravitational sector consists only of gravitons. In terms of bulk quantities, the central charge is 
\beq
c_T=-\frac{1}{4}\frac{\Mp^2}{H^2}\ .
\eeq
In a quasi-de Sitter background, with finite slow-roll parameter $\varepsilon$, some of the conformal symmetries are softly broken (see Appendix~\ref{app:cftbreaking}). 
The effects of this weak symmetry breaking can be treated perturbatively.

\subsection{Conformal Perturbation Theory}
\label{sec:CPT}

A suitable framework for analyzing field theories that are almost conformal is conformal perturbation theory~\cite{Zamolodchikov:1987ti}. 
We now wish to show that such an analysis reproduces the results of Section~\ref{sec:bulk}.

\vskip 4pt
We deform the CFT by a local primary operator~\cite{van2004inflationary}\hskip 1pt\footnote{Of course, CFTs are characterized by a set of correlation functions rather than by an action. Here, $S_{\mathsmaller{\rm CFT}}$ is simply a metaphoric way of characterizing the content of the original CFT. In practice, calculations in conformal perturbation theory are always performed at the level of correlation functions.} 
\beq
S = S_{\mathsmaller{\rm CFT}} + \varphi \int \dd^3 z \,  {\cal O}(\z \hskip 1pt)\ ,
\eeq
where $\varphi$ is a small coupling.\footnote{In conformal perturbation theory, one usually tunes the perturbation so that the beta function vanishes and the theory flows to a new conformal fixed point~\cite{Zamolodchikov:1987ti}. Since we are mainly interested in the parametric scaling of the corrections, we will not perform this additional step and thus we do not worry about the particular renormalization group flow.} We take the perturbing operator to be marginally relevant, so its dimension is $\Delta \equiv 3 -\lambda$, with $0< \lambda \ll 1$. The small expansion parameter dual to $\sqrt{\varepsilon}$ will be a combination of $\varphi$ and $\lambda$.
For convenience, we normalize the two-point function of ${\cal O}$ by the central charge
\beq
\langle {\cal O}_\k \hskip 1pt {\cal O}_{-\k} \rangle' = c_T\hskip 1pt k^{3- 2\lambda}\ .
\eeq
For small $\varphi$, the two-point function of the stress tensor  
can be computed perturbatively as
\begin{align}
\langle T_{ij} T_{kl} \rangle &= \big\langle T_{ij} T_{kl} \,e^{-\varphi\int \d^3 z \, {\cal O}} \big\rangle_{0}  \nonumber \\[10pt]
&= \big\langle T_{ij} T_{kl} \big \rangle_{0}  - \varphi \int \d^3 z\, \big\langle T_{ij} T_{kl}\, {\cal O}(\z\hskip 1pt)\big\rangle_{0}  \nonumber  \\[4pt]
& \hspace{2.02cm}+\frac{\varphi^2}{2}\int \dd^3z\,\dd^3w\,  \big\langle T_{ij} T_{kl}\,{\cal O}(\vec{z}\hskip 1pt) {\cal O}(\vec{w})\big\rangle_{0}+\cdots \  , \label{equ:expansion}
\end{align}
where the expectation values $\langle T_{ij}T_{kl} \ldots {\cal O}\rangle_0$ are computed using the CFT operator algebra, and in general are constrained by Ward identities.  We will use the following trace Ward identities obeyed by the stress tensor 
\begin{align}
\big\langle T^i{}_i(\x \hskip 1pt)\hskip 1pt T_{kl}(\y \hskip 1pt)\hskip 1pt {\cal O}(\z \hskip 1pt) \big\rangle_{0} &= \lambda\, \delta(\x-\z \hskip 1pt)\, \big\langle T_{kl}(\y \hskip 1pt)\hskip 1pt {\cal O}(\z \hskip 1pt) \big\rangle_{0} =0\ , \label{equ:trace1}\\[10pt]
\big\langle T^i{}_i(\x \hskip 1pt)\hskip 1pt T_{kl}(\y \hskip 1pt)\hskip 1pt {\cal O}(\z \hskip 1pt){\cal O}(\vec{w}) \big\rangle_{0} &= \lambda  \left[\delta(\x-\z \hskip 1pt)\, \big\langle T_{kl}(\y \hskip 1pt) \hskip 1pt {\cal O}(\z \hskip 1pt){\cal O}(\vec{w})\big\rangle_{0} + (\z \leftrightarrow \vec w) \right]\ , \label{equ:trace2}
\\[10pt]
\big\langle T^i{}_i(\x \hskip 1pt)\hskip 1pt T^k{}_{k}(\y \hskip 1pt)\hskip 1pt {\cal O}(\z \hskip 1pt){\cal O}(\vec{w}) \big\rangle_{0} &= \lambda^2  \left[\delta(\x-\z \hskip 1pt)\, \delta(\y-\vec{w}) \big\langle \hskip 1pt {\cal O}(\z \hskip 1pt){\cal O}(\vec{w})\big\rangle_{0} + (\z \leftrightarrow \vec w) \right]\  , \label{equ:trace3}
\end{align}
where, in the last identity, we have dropped an irrelevant contact term, 
with support when $\x=\y$.  In a reparametrization invariant theory,  $\nabla^i\langle T_{ij} \rangle=0$, we furthermore have $k^i\langle T_{ij}(\k) T_{kl}(-\k) \rangle' =0$. Imposing this constraint implies that $\langle T_{ij} T_{kl}\rangle$ has the following form
\beq
\langle T_{ij}(\k\hskip 1pt) T_{kl}(-\k\hskip 1pt) \rangle' = {1\over 4}\left[\delta^\perp_{ij}\delta^\perp_{kl} \langle T_{\k} T_{-\k}\rangle'+ \left(\delta^\perp_{ik}\delta^\perp_{jl}+\delta^\perp_{il}\delta^\perp_{jk}-\delta^\perp_{ij}\delta^\perp_{kl}\right) \langle T^s_{\k}  \hskip 1pt T^s_{-\k}\rangle' \right]\ , \label{equ:TT}
\eeq
where $\delta^\perp_{ij} \equiv \delta_{ij}-{k_i k_j/ k^2}$. 

\vskip 4pt
In general, $\langle T^s_{\k}  \hskip 1pt T^s_{-\k}\rangle'$ and $ \langle T_{\k}  \hskip 1pt T_{-\k}\rangle'$ are arbitrary functions of $k$, but in a theory with approximate conformal symmetry, we expect  them to be approximately scale invariant.  The breaking of scale invariance can be studied in powers of $\varphi$:
\begin{itemize}
\item First, let us look at the two-point function of the trace, $\langle T_{\k} \hskip 1ptT_{-\k}\rangle$. It follows from the trace Ward identity (\ref{equ:trace1}) that this vanishes at order $\varphi$.  The scalar two-point function is therefore only generated at order $\varphi^2$. Using (\ref{equ:trace2}) and (\ref{equ:trace3}), we find  
\beq
\langle T_{\k} \hskip 1ptT_{-\k}\rangle' =  \varphi^2 \lambda^2 \langle {\cal O}_{\k} {\cal O}_{-\k}\rangle'\,=\, c_T\, \varphi^2 \lambda^2 \, k^{3- 2 \lambda}\ .
\eeq
The tensor-to-scalar ratio therefore is 
\beq
r \equiv \frac{\langle T_{\k}  \hskip 1pt T_{-\k} \rangle'}{\langle T_{\k}^s  \hskip 1pt T_{-\k}^s\rangle'}  = \varphi^2 \lambda^2 \ .
\eeq
Comparing this to the bulk result, $r = 16\varepsilon$, we identify the following duality map: $\varphi \lambda \leftrightarrow \pm 4 \sqrt{\varepsilon}$.

\item Next, we consider the correction to $\langle T^s_{\k}  \hskip 1pt T^s_{-\k}\rangle'$. At $O(\varphi)$, we require the integral of the three-point function $\langle T^s T^s {\cal O}\rangle_0$. 
In position space, we have
\beq
\big \langle T_{ij}(\x\hskip 1pt) T_{kl}(\y\hskip 1pt) {\cal O}(\z\hskip 1pt)\big \rangle'_0= c_T f_{TT\cal O} \, {\cal T}_{ij\hskip 1pt kl}(\x-\y\hskip 1pt,\hskip 1pt\y-\z\hskip 1pt,\hskip 1pt\z-\x\hskip 1pt)\ , \label{equ:315}
\eeq
where an explicit expression for the tensor structure ${\cal T}_{ij \hskip1pt kl}$ can be found in the classic work of Osborn and Petkou~\cite{Osborn:1993cr}, cf.~eqs.~(3.3)--(3.6).
We have identified $c_T f_{TT\cal O}$ with the coefficient~$a$ of eq.~(3.4) in~\cite{Osborn:1993cr}.
Integrating (\ref{equ:315}) over $\z$, and transforming to momentum space, we get
\beq
 \big\langle T^s_{\k} \hskip 1pt T^s_{-\k} \,{\cal O}(\vec 0\hskip 1pt) \big\rangle_{0}' = c_T\hskip 1pt\alpha(\lambda)  f_{TT{\cal O}}\, k^{3 - \lambda}\ , \label{equ:TTO}
\eeq
where $\alpha(\lambda)$ is a numerical coefficient. 
 We have confirmed by explicit integration that $\alpha(\lambda)$ is finite, even in the limit $\lambda \to 0$.\footnote{We also found that $\alpha(\lambda)$ vanishes for a two-dimensional CFT.  This is to be expected from the $c$-theorem in two dimensions. It is also consistent with our bulk interpretation, since the Weyl tensor vanishes identically in three dimensions.}  This implies that $T_{ij}$ doesn't need to be renormalized, and also means that perturbing the CFT by an exactly marginal operator simply shifts the coefficient of the stress tensor two-point function.  Substituting (\ref{equ:TTO}) into (\ref{equ:expansion}), we get
\beq
\langle T_{\k}^s  \hskip 1pt T_{-\k}^s\rangle' =  c_Tk^3  \big(1- n_t \ln k + \cdots\big) \  ,
\eeq
where we have dropped a small $O(\varphi)$ shift of the amplitude, and defined 
\beq
n_t \equiv - \varphi \lambda\, \alpha(\lambda)\, f_{TT{\cal O}}\ .
\eeq
This reproduces the $O(\sqrt{\varepsilon})$ contribution in the bulk result (\ref{equ:nt}), if we make the following identification: 
$\alpha(\lambda) f_{TT{\cal O}} \leftrightarrow \sqrt{2}  \hskip 1pt b \hskip 1pt H^2/M^2$.  The $O(\varphi^2)$ term in the $\langle T^s_\k  \hskip 1pt T^s_{-\k}\rangle$ correlator contains the standard tensor tilt proportional to $\varepsilon$. This contribution depends on the details of the CFT and its various OPE coefficients. 
\end{itemize}
The main result of this section was the confirmation that a tensor tilt is generated at $O(\varphi)$, while the tensor-to-scalar ratio is only non-zero at $O(\varphi^2)$. The tensor tilt comes from a specific three-point function, whose size is set by $f_{TT\cal O}$ in the boundary CFT, and by $H^2/M^2$ in the bulk action. The standard result of Einstein gravity is recovered for $f_{TT\cal O} \to 0$, or $M \to \infty$.

\section{Conclusions}
\label{sec:conclusions}

The weak breaking of conformal symmetry during inflation can be used to constrain the predictions for cosmological correlators---e.g.~\cite{Maldacena:2011nz,Mata:2012bx,Arkani-Hamed:2015bza}.  This is especially relevant in inflationary models in which the scale suppressing higher-curvature corrections is close to the Hubble scale.  In this paper, we have studied the coupling of the inflaton field to higher-curvature tensors in models with a minimal breaking of conformal symmetry. We showed that the most general correction to the tensor two-point function is captured by a coupling to the square of the Weyl tensor. This interaction modifies the consistency condition of single-field slow-roll inflation
\beq
- \frac{8 \hskip 1pt n_t}{r} \,=\, 1 \mp \frac{4 b}{\sqrt{2\varepsilon}} \frac{H^2}{M^2}\ .
\eeq
The correction can have either sign, and may dominate over the prediction from Einstein gravity if $H/M$ is not too small. We consider this an interesting signature of higher-curvature corrections during inflation. 

\vskip 10pt
We have left a few open questions for future work:
\begin{itemize}
\item What is the precise connection between our effective action  \eqref{eq:action}  and the original Starobinsky model~\cite{Starobinsky:1980te}? Both models rely on softly broken conformal invariance, and the effective actions even contain some terms of the same functional form. Making this relationship more precise would be very interesting~\cite{future}.

\item How naturally does our scenario arise in explicit string compactifications?
Under which circumstances is a weakly broken conformal symmetry maintained in the four-dimensional effective theory? Is there a relation to conformal supergravity~\cite{fradkin1985conformal, berkovits2004conformal,maldacena2011einstein}?

\item If a violation of the tensor consistency condition were to be observed, how would we convince ourselves that it comes from higher-curvature effects? In particular, it is well-known that a violation of $r=-8n_t$ can also arise from modifications of the scalar spectrum in models with a nontrivial scalar sound speed~\cite{Chen:2006nt} and/or isocurvature fluctuations~\cite{wands2002observational}. 
However, in that case we also expect strong interactions in the scalar sector, which can be tested for through measurements of primordial non-Gaussianity. 
In contrast, in our proposal we do not predict a strong counterpart in scalar non-Gaussianity.
A positive test of our scenario would be looking for correlated signatures of a low string scale, such as angular dependence in the scalar bispectrum~\cite{Arkani-Hamed:2015bza} and specific forms of tensor non-Gaussianity~\cite{Maldacena:2011nz}.

\item Can we get blue tensors? Our analysis determines neither the sign of the coupling to the Weyl tensor, nor its time dependence. While the sign may be constrained by requiring  tensors to propagate subluminally, we see no a priori way to constrain the rate of change of the coupling.
At the moment, blue tensors therefore seem to be a legitimate possibility.
\end{itemize}

\vspace{0.5cm}
\subsubsection*{Acknowledgements}

We thank Mustafa Amin, David Berenstein, Paolo Creminelli, Garrett Goon,  Liam McAllister, Paul McFadden, Mehrdad Mirbabayi, Jorge Nore\~na, Rafael Porto, Jan Pieter van der Schaar,  Ilya Shapiro, Stephen Shenker, Filippo Vernizzi, and especially Juan Maldacena and Hugh Osborn for helpful discussions. We thank Juan Maldacena, Liam McAllister and Mehrdad Mirbabayi for comments on a draft.
G.P.~thanks the KITP for hospitality during the program `Quantum Gravity Foundations: UV to IR'. Research at the KITP is supported in part by the National Science Foundation under Grant No. NSF PHY11-25915.
D.B.~and G.P.~thank the Aspen Center for Physics (NSF Grant 1066293) for their hospitality while this work was being completed. 
D.B., H.L.~and G.P.~acknowledge support from a Starting Grant of the European Research Council (ERC STG Grant 279617). H.L.~is supported by the Cambridge Overseas Trust and the William Georgetti Scholarship.

\newpage
\appendix
\section{Comments on Conformal Symmetry}
\label{app:cftbreaking}

In this appendix, we study the breaking of conformal symmetry in an inflationary spacetime. By conformal symmetry, we mean the action of the de Sitter isometries on quantum fields in the background geometry. We will show that dilatations and special conformal transformations are broken by an amount controlled by $\varepsilon = - \dot H/H^2$.
Moreover, we will demonstrate that in theories with nontrivial sound speed, $c_s \ll 1$, special conformal symmetry is broken even in the limit~$\varepsilon\to0$.

\begin{itemize}
\item First, let us consider a massless scalar field $f$ in pure de Sitter space
\beq\label{eq:acscf}
S_{dS} ={1\over2}\int \dd^4x\,\frac{(f')^2-(\vec{\nabla} f)^2}{(H\eta)^2}\ .
\eeq
Conformal transformations act as follows 
\begin{align}
\delta_\lambda f &= \lambda \left(\eta f' + \x\cdot\vec\nabla f\right) \, , \label{eq:d} \\
\delta_b f &=2\hskip 1pt \b\cdot \x \left(\eta f'+ \x\cdot\vec\nabla f\right)+(\eta^2-x^2)\,\b\cdot\vec\nabla f\ . \label{eq:sct}
\end{align}
It is straightforward to check that the action \eqref{eq:acscf} is invariant under these transformations. 

\item Next, let us repeat the analysis for an inflationary background. The action of a massless scalar in quasi-de Sitter space is
\beq
S_I = {1\over2}\int \dd^4x~a^2(\eta)\left((f')^2-(\vec{\nabla} f)^2\right) \equiv \int \dd^4 x\, {\cal L}_I \ , \qquad a(\eta)=-\frac{1}{H\eta(1-\varepsilon)}\ .
\eeq

After  integrations by parts, we obtain the following variations of the action
\begin{align}
\delta_\lambda S_I&= 2 \lambda \int \dd^4 x\, \varepsilon\, {\cal L}_{I} \ ,\\
\delta_b S_I &=  2 \int \dd^4 x\, (\b\cdot \x \hskip 1pt)\, \varepsilon\, {\cal L}_{I} \ .
\end{align}
As advertised, dilatations and SCTs are broken by an amount proportional to $\varepsilon$.
 
\item Finally, we consider a massless scalar field with a nontrivial speed of sound
\beq\label{eq:csaction}
S_{c_s} ={1\over2}\int \dd^4x\, \frac{1}{c_s^2}\frac{(f')^2- c_s^2(\vec{\nabla} f)^2}{ (H\eta)^2}\ .
\eeq
Assuming $c_s = const.$, for simplicity, the variation of the action gives
\begin{align}
\delta_\lambda S_{c_s}&= 0 \ ,\\
\delta_b S_{c_s} &= 2 \int \dd^4 x\, \frac{1-c_s^2}{c_s^2}\, \frac{\eta f'\, (\b\cdot\vec{\nabla})f}{ (H\eta)^2} \ .
\end{align}
 
We see that scale invariance is retained, while special conformal invariance is broken. 
For~time-dependent $c_s(t)$, dilatations would be broken as well. 

\end{itemize}

 \newpage
\section{Comments on Einstein Frame}
\label{app:Einstein}
 
In \cite{Creminelli:2014wna}, it was shown that a nontrivial tensor sound speed can be set to unity by a disformal transformation~\cite{Bekenstein:1992pj}. This is followed by a conformal transformation, which brings the action back to Einstein frame. The combined transformation is given by
\beq
g_{\mu\nu} \ \to\ c_t^{-1}\left[g_{\mu\nu}+(1-c_t^2)\hskip 1pt n_\mu n_\nu\right]\, , \label{equ:disformal}
\eeq
where $n_\mu \propto \partial_\mu \phi$ is the unit vector orthogonal to the constant-time hypersurfaces.
The action in the new frame then has a trivial sound speed for tensors, $\tilde c_t = 1$, but a nontrivial sound speed for scalars, $\tilde c_s = c_t^{-1}$.  In this appendix, we show that observables are the same in both frames. In particular, we will find that the modification to the consistency condition (\ref{eq:consistcond}) is still present in the new frame.

\vskip 4pt
Consider the action (\ref{eq:action}) in comoving gauge.
At quadratic order in fluctuations and at leading order in slow-roll, we get
\beq
S=\frac{\Mp^2}{8}\int \d t\hskip 1pt\d^3x\, \big({\cal L}_\zeta+{\cal L}_\gamma\big)\ , \quad \ \text{with} \quad \ \begin{aligned}{\cal L}_\zeta &\, = \, 8\hskip 1pt a^3\hskip 1pt\varepsilon \left[\dot\zeta^2-a^{-2}(\vec\nabla\zeta)^2 \right] \, ,\\[4pt]
 {\cal L}_\gamma &\, =\,  a^3\hskip 1pt c_t^{-2}\left[\dot\gamma_{ij}^2 - a^{-2}c_t^2(\vec\nabla\gamma_{ij})^2 \right]\, .\label{eq:comoving}
 \end{aligned}
\eeq
After performing the transformation (\ref{equ:disformal}), the background line element becomes $\d s^2=-c_t\hskip 1pt \d t^2+c_t^{-1}a^2\d\x\hskip 1pt{}^2$.
 Rescaling the time and the scale factor,
\beq
\d\tilde t = c_t^{1/2}(t)\hskip 2pt\d t \ , \quad\ \tilde a(\tilde t\hskip 1pt)=c_t^{-1/2}(t)\hskip 2pt a(t) \ ,
\eeq
we get $\d s^2=-  \d \tilde t\hskip 1pt{}^2+ \tilde a^2\d\x\hskip 1pt{}^2$.  The curvature perturbation $\zeta$ and the tensor fluctuations $\gamma_{ij}$ transform as spacetime scalars, so 
the action (\ref{eq:comoving}) takes the form
\beq
S=\frac{\Mp^2}{8}\int \d \tilde t  \hskip 1pt  \d^3x\, \big(\tilde{\cal L}_{\tilde \zeta}+\tilde{\cal L}_{\tilde \gamma}\big)\ , \quad \ \text{with} \quad \ \begin{aligned}\tilde{\cal L}_{\tilde \zeta} &\, = \, 8 \hskip 1pt \tilde a^3\hskip 1pt \varepsilon \left[ c_t^2 (\partial_{\tilde t}\hskip 1pt \tilde \zeta)^2-\tilde a^{-2}(\vec\nabla\tilde\zeta)^2 \right] \, ,\\[4pt]
 \tilde{\cal L}_{\tilde \gamma} &\, =\,  \tilde a^3 \left[(\partial_{\tilde t}\hskip 1pt \tilde\gamma_{ij})^2 - \tilde a^{-2}(\vec\nabla\tilde\gamma_{ij})^2 \right]\, .
 \end{aligned}\label{eq:comovingtilde}
\eeq
Hence, in the new frame, the tensors propagate with a trivial sound speed, 
$\tilde c_t = 1$, but the scalars have a modified 
sound speed, $\tilde c_s = c_t^{-1}$. Notice that $c_t < 1$ implies $\tilde c_s > 1$.
It is not unusual that a non-local field redefinition maps a purely luminal theory to one with apparent superluminality (e.g.~\cite{Creminelli:2013fxa, deRham:2013hsa, deRham:2014lqa}).  In such a situation, the presence of a superluminal mode does not 
 imply a violation of relativistic causality.\footnote{We thank Paolo Creminelli for explaining this to us.}

\vskip 4pt
At leading order in slow-roll, the tensor power spectrum takes the standard form
\beq
 \Delta_{\gamma}^2 = \frac{2}{\pi^2}\frac{\tilde H^2}{\Mp^2}\ ,
\eeq
in terms of the new Hubble parameter $\tilde H \equiv \partial_{\tilde t} \ln \tilde a \approx c_t^{-1/2}H$.  The tensor tilt is hence also of the usual form, $n_t = -2 \tilde \varepsilon$, and all nontrivial features have been moved to the scalar sector. The power spectrum of curvature perturbations is 
\beq
\Delta_{\zeta}^2 = \frac{1}{8\pi^2}\frac{1}{\varepsilon \tilde c_s}\frac{\tilde H^2}{\Mp^2}\ , 
\eeq
where
\beq
\varepsilon = \tilde\varepsilon+\frac{1}{2}\tilde\varepsilon_s\ , \quad \text{with} \quad \tilde\varepsilon_s = -\varepsilon_t\ .
\eeq
 If we neglect the small shift in the amplitude due to $\tilde c_s \approx 1$, then the tensor-to-scalar ratio is
\beq
r \approx  16 \left[\tilde\varepsilon+\frac{1}{2}\tilde\varepsilon_s\right] \ .
\eeq
Hence, although the tensor tilt is standard in the new frame, the tensor-to-scalar ratio now is non-standard.
The tensor consistency condition is then given by
\beq
-\frac{8n_t}{r}\ =\ \frac{\tilde\varepsilon}{\tilde\varepsilon+\frac{1}{2}\tilde\varepsilon_s}\ \left(= 1+\frac{1}{2}\frac{\varepsilon_t}{\varepsilon}\right)\, .
\eeq
We see that the consistency condition is still modified in the new frame, but now the effect is  coming from the time dependence of a nontrivial scalar sound speed, $\tilde \varepsilon_s \ne 1$. 
Substituting $\tilde\varepsilon$ and $\tilde\varepsilon_s$ in terms of the parameters in the original frame, $\varepsilon$ and $\varepsilon_t$, we find complete agreement with our previous result (\ref{eq:consistcond}).

\newpage
\addcontentsline{toc}{section}{References}
\bibliographystyle{utphys}
\bibliography{Refs}

\end{document}